\documentclass[twocolumn,iop,twocolappendix]{emulateapj}

\usepackage{amssymb,amsmath}
\usepackage{mathrsfs}
\usepackage{times}
\usepackage{bm}

\newcommand{\beq}{\begin{equation}}
\newcommand{\eeq}{\end{equation}}
\newcommand{\intd}{{\rm d}}
\newcommand{\msun}{\ensuremath{M_\odot}}
\newcommand{\tp}{\ensuremath{t_{\rm P}}}

\newcommand{\texp}{\ensuremath{t_{0}}}

\newcommand{\bol}{\ensuremath{_{\rm bol}}}
\newcommand{\lbol}{\ensuremath{L\bol}}

\newcommand{\nif}{\ensuremath{^{56}}Ni}
\newcommand{\mni}{\ensuremath{M_{\rm Ni}}}

\newcommand{\eexp}{\ensuremath{E_{\rm exp}}}
\newcommand{\mej}{\ensuremath{M_{\rm ej}}}
\newcommand{\lpl}{\ensuremath{L_{\rm pl}}}
\newcommand{\tpl}{\ensuremath{\Delta t_{\rm pl}}}
\newcommand{\tni}{\ensuremath{\Delta t_{\rm Ni}}}
\newcommand{\cf}{\ensuremath{\mathbf{C}^{\mathbf{f}}}}
\newcommand{\ca}{\ensuremath{\mathbf{C}^{\mathbf{a}}}}
\newcommand{\ag}{\ensuremath{A_\gamma}}

\defcitealias{pp14}{PP15}

\begin{document}
\title{On The Intrinsic Diversity of Type II-Plateau Supernovae}
\shorttitle{On The Intrinsic Diversity of Type II-Plateau Supernovae}
\shortauthors{Pejcha \& Prieto}

\author{ Ond\v{r}ej Pejcha\altaffilmark{1}}
\affil{Department of Astrophysical Sciences, Princeton University, 4 Ivy Lane, Princeton, NJ 08540, USA}
\email{pejcha@astro.princeton.edu}
\and
\author{Jose L. Prieto}
\affil{N\'ucleo de Astronom\'ia de la Facultad de Ingenier\'ia, Universidad Diego Portales, Av. Ej\'ercito 441, Santiago, Chile}
\affil{Millennium Institute of Astrophysics, Santiago, Chile}
\altaffiltext{1}{Hubble and Lyman Spitzer Jr.\ Fellow}

\begin{abstract}
Hydrogen-rich Type II-Plateau supernovae exhibit correlations between the plateau luminosity $\lpl$, the nickel mass $\mni$, the explosion energy $\eexp$, and the ejecta mass $\mej$. Using our global, self-consistent, multi-band model of nearby well-observed supernovae, we find that the covariances of these quantities are strong and that the confidence ellipsoids are oriented in the direction of the correlations, which reduces their significance. By proper treatment of the covariance matrix of the model, we discover a significant intrinsic width to the correlations between $\lpl$, $\eexp$ and $\mni$, where the uncertainties due to the distance and the extinction dominate. For fixed $\eexp$, the spread in $\mni$ is about $0.25$\,dex, which we attribute to the differences in the progenitor internal structure. We argue that the effects of incomplete $\gamma$-ray trapping are not important in our sample. Similarly, the physics of the Type II-Plateau supernova light curves leads to inherently degenerate estimates of $\eexp$ and $\mej$, which makes their observed correlation weak. Ignoring the covariances of supernova parameters or the intrinsic width of the correlations causes significant biases in the slopes of the fitted relations. Our results imply that Type II-Plateau supernova explosions are not described by a single physical parameter or a simple one-dimensional trajectory through the parameter space, but instead reflect the diversity of the core and surface properties of their progenitors. We discuss the implications for the physics of the explosion mechanism and possible future observational constraints.
\end{abstract}

\keywords{Methods: statistical --- stars: distances --- supernovae: general}

\section{Introduction}

The observed light curves and expansion velocities of hydrogen-rich Type II-Plateau supernovae can be used to infer the properties of the explosions and progenitor stars. Specifically, the duration and luminosity of the optically-thick ``plateau'' phase of nearly constant bolometric luminosity is primarily set by the explosion energy $\eexp$ and ejecta mass $\mej$ \citep[e.g.][]{arnett80,kasen09}. The subsequent nearly exponential fading is powered by the thermalization of radioactive fission products of \nif\ and the luminosity is thus proportional to the nickel mass $\mni$ \citep[e.g.][]{hamuy03}. Patterns in the distributions and the correlations between these quantities can guide the stellar evolution and explosion models, where many open questions persist \citep[e.g.][]{burrows13,ugliano12,pt12,pt14,pejcha_ns,pejcha_cno,prieto08a,prieto08b,prieto08c,prieto12,prieto13,holoien14,ertl15}. For example, it has been proposed that the supernova explosion energy is proportional to the ejecta mass and therefore also the progenitor mass \citep[e.g.][]{hamuy03,utrobin09,poznanski13}. However, low-energy explosions might be an exception signaling significant fallback in massive progenitors \citep[e.g.][]{zampieri03,pastorello04,nomoto06}. At the same time, direct progenitor detections show significant scatter but little correlation between the progenitor mass and $\mni$ \citep{smartt09}.

Naturally, not all supernova parameters can be inferred independently. For example, quantities based on bolometric luminosity such as $\mni$ or the plateau luminosity $\lpl$ are plagued by systematic uncertainties in the distance or extinction. Similarly, a simultaneous change in several parameters can result in nearly identical light curves \citep{arnett80,woosley88,popov93,kasen09,dessart10,nagy14}. 

Here, we evaluate whether or not the systematic uncertainties and parameter covariances influence the significance of the correlations between parameter estimates for Type II-Plateau supernovae. We focus on $\lpl$, $\mni$, $\eexp$, and $\mej$. To this end, we employ the self-consistent global model of nearby well-observed Type II-Plateau supernovae that we developed in \citet[hereafter PP15]{pp14}, which simultaneously fits multi-band light curves and expansion velocities and provides consistent distances, reddenings, bolometric luminosities and, most importantly for the present purposes, their covariances. In Section~\ref{sec:par}, we describe how we estimate the supernova parameters, the database of supernova observations and the estimates of uncertainties. In Section~\ref{sec:scatter}, we investigate the systematic uncertainty due to distance and discover a significant intrinsic width of the $\eexp$--$\mni$ correlation. In Section~\ref{sec:m_eexp}, we address the significance of the $\mej$--$\eexp$ correlation. In Section~\ref{sec:disc}, we discuss the astrophysical implications of our findings.

\section{Supernova parameters derived from observations}
\label{sec:par}

We calculate the bolometric luminosity of the optically-thick plateau phase $\lpl$ as 
\beq
\lpl = \lbol(\texp + \tpl),
\label{eq:lpl}
\eeq
where $\lbol$ is the bolometric light curve obtained by integrating the spectral energy distribution from about $0.19$ to $2.2\,\mu$m and extrapolating the Rayleigh-Jeans tail \citepalias{pp14}, and $\texp$ is the zero point of our model fits, which coincides with the explosion epoch for the purposes of this paper. The luminosity is evaluated at a fixed interval $\tpl$ after the explosion epoch $\texp$. Following \citet{hamuy03}, our fiducial choice is $\tpl = 50$\,days, but we will investigate the sensitivity of some of our results to $\tpl$. We estimate the nickel mass $\mni$ from the exponential decay tail of the light curve after \citet{hamuy03} as
\beq
\mni = 7.866\times 10^{-44} \lbol(\texp+\tni)\exp\left(\frac{\tni-6.1\,{\rm d}}{\tau}\right) \msun,
\label{eq:mni}
\eeq
where $\tau = 111.26$\,days. Our fiducial choice for the time elapsed after explosion, where we estimate $\mni$, is $\tni = 200$\,days. The units on $\lbol$ are ergs\,s$^{-1}$.

We estimate the explosion energy $\eexp$ and ejected mass $\mej$ using linear relations of the form
\begin{eqnarray}
\log \left(\frac{\eexp}{10^{50}\,{\rm ergs}}\right) & = & \bm{\alpha} \cdot \mathbf{b} + \eta_{\rm exp}, \label{eq:eexp}\\
\log \left(\frac{\mej}{\msun}\right) & = & \bm{\beta} \cdot \mathbf{b} + \eta_{\rm ej}, \label{eq:mej}
\end{eqnarray}
where $\mathbf{b} = (M_V, \log \tp, \log v)$, and $M_V$ is the absolute magnitude in the $V$ band, $\tp$ is the duration of the optically-thick plateau phase measured at the midpoint of the drop to the exponential decay phase\footnote{The plateau duration $\tp$ should not be confused with the time, when the plateau luminosity is measured, $\tpl$.}, and $v$ is the expansion velocity of the photosphere in the units of $1000$\,km\,s$^{-1}$ commonly measured on the Fe II 5169\,\AA\ line. $M_V$ and $v$ are evaluated at the midpoint of the plateau, corresponding to time $\texp + \tp/2$. The coefficient vectors $\bm{\alpha}$ and $\bm{\beta}$ are typically obtained either from analytic models of supernova light curves \citep[e.g.][]{arnett80,popov93,kasen09} or from the fits to the simulated light curves and expansion velocities \citep[e.g.][]{litvinova83,litvinova85,kasen09}. Here, we use coefficients from the analytic model of \citet{popov93}, $\bm{\alpha} = (0.4,4.0,5.0)$, $\bm{\beta} = (0.4, 4.0, 3.0)$, $\eta_{\rm exp}=-3.311$, and $\eta_{\rm ej} = -2.089$, and the simulations of \citet{litvinova85}, $\bm{\alpha} = (0.135,2.34,3.13)$, $\bm{\beta} = (0.234,2.91,1.96)$, $\eta_{\rm exp}=-3.205$, and $\eta_{\rm ej} = -1.829$. \citet{litvinova85} claim that Equations~(\ref{eq:eexp}--\ref{eq:mej}) can reproduce their numerical results to about $30\%$ and their results have been commonly used in the literature \citep[e.g.][]{elmhamdi03,hamuy03,hendry06,bose13}. We emphasize that the point of this exercise is not to compete with detailed modeling employing sophisticated codes, but to illustrate the limitations posed by the physics of the supernova light curves to the estimates of $\eexp$ and $\mej$ -- the analytic results for $\bm{\alpha}$ and $\bm{\beta}$ of \citet{popov93} already imply that the estimates of $\eexp$ and $\mej$ are highly correlated. We will show in Section~\ref{sec:m_eexp} that similar degeneracy is present also in the more sophisticated models.

To estimate the above parameters, we use the model and database of observations from \citetalias{pp14}. The light curves are described by a set of phenomenological parameters $\mathbf{a}$ obtained by least-squares fitting of the data, which also provides the full covariance matrix $\ca$ of our model. This allows us to properly propagate uncertainties in $\mathbf{a}$ to $\lpl$, $\mni$, $\eexp$, and $\mej$. In this work, we employ a model fit with observational uncertainties rescaled so that the final reduced $\chi^2$ is unity, which increases the values in $\ca$ relative to the unadjusted fit\footnote{We repeated the analysis with unadjusted covariance matrix and found that line slopes, intercepts and their uncertainties remain unchanged. The intrinsic widths are about $20\%$ higher when using the unadjusted full covariance matrix. The Bayes factors $B$ change by only about $20\%$, but $R$ increases in all cases by a factor of $\sim 2.5$. Finally, the unadjusted covariance matrix will yield smaller confidence ellipsoids in the Figures.}. We use a subset of the observational sample of \citetalias{pp14} that does not include SN2007od, SN2006bp, and SN2002hh due to reasons mentioned in \citetalias{pp14}. In addition, we include a recent Type II-P SN2013am \citep{zhang14}, which we fit using the publicly available version of our fitting tool \footnote{\url{http://www.astro.princeton.edu/~pejcha/iip/}}. For SN2013am, we find an explosion epoch $\texp = 2456373.0 \pm 2.4$, total reddening $E(B-V) = 0.81 \pm 0.02$, and a distance modulus $29.2 \pm 0.3$\,mag. For every choice of $\tpl$ and $\tni$ we select a subset of supernovae with data before and after these dates to prevent extrapolation of the model. Our fiducial choice of $\tpl$ and $\tni$ thus leaves us with $19$ supernovae, which satisfy these constraints.

We calculate the covariance matrix $\cf$ of $\mathbf{f} = \log (\lpl, \mni, \eexp, \mej)$  using the standard procedure for uncertainty propagation 
\beq
\cf = \left(\frac{\partial \mathbf{f}}{\partial\mathbf{a}}\right)\ca \left(\frac{\partial \mathbf{f}}{\partial\mathbf{a}}\right)^{\rm T}.
\label{eq:propag}
\eeq
The confidence ellipsoid for $\bm{f}$ is a quadratic equation in the offsets $\delta \mathbf{f}$
\beq
\Delta\chi^2 = \delta\mathbf{f} (\cf)^{-1} \delta\mathbf{f}^{\rm T},
\label{eq:ellipse}
\eeq
where $\Delta\chi^2$ depends on the desired confidence level and the number of variables. In the subsequent discussion, we will exclusively focus on pairs drawn from $\mathbf{f}$, $(f_i, f_j)$, their $2\times 2$ covariance submatrix $\mathbf{C}^{f_i,f_j}$, and the $68.3\%$ confidence level, where $\Delta\chi^2 \approx 2.30$ \citep[p.~697]{press92}. For two parameters at a time, Equation~(\ref{eq:ellipse}) can be solved to obtain the positions on the confidence ellipsoid by transforming the pair of variables $\delta\mathbf{f}=(\delta f_i, \delta f_j)$ to polar coordinates, varying the polar angle, and solving for the radial distance at each polar angle. Note that the off-diagonal elements of $\cf$ can be non-zero even if $\ca$ is diagonal. In our covariance matrix, we properly take into account the uncertainties in $\texp$ and their associated covariances with the distance and plateau duration.

We model the dependencies between individual components of $\mathbf{f} = \log (\lpl, \mni, \eexp, \mej)$ with a straight line allowing for an intrinsic width $\Sigma$. We use the likelihood function of \citet{hogg10}, which assumes that the observations are offset from a linear relation by a Gaussian described by the covariance matrix $\mathbf{C}^{f_i,f_j}$ convolved with a Gaussian intrinsic scatter with standard deviation $\Sigma$, which is perpendicular to the linear relationship. We obtain the confidence intervals with the MCMC sampler {\tt emcee} \citep{foreman13}. We will quote median of the distribution for our best-fit parameters, and $16$ and $84$ percentiles as their confidence intervals.

We quantify the significance of the correlations in the data using two approaches. First, we calculate the Bayes factor $B$ of the linear fit relative to a model, which assumes no correlation between the two variables. Specifically, we evaluate
\beq
B \equiv \frac{\frac{1}{\pi}\int_{-\pi/2}^{\pi/2} P(\theta)\intd \theta}{\max[P(\theta=0),P(\theta = \pi/2)]},
\label{eq:bayes}
\eeq
where $\theta$ is the angle between the line and the $x$ axis and $P(\theta)$ is the likelihood marginalized over the line intercept and intrinsic width together with their priors; we assume flat prior in $\theta$.  Taking the maximum of $P(\theta=0)$ and $P(\theta=\pi/2)$ in the denominator of Equation~(\ref{eq:bayes}) ensures that $B \approx 1$ for uncorrelated data when the scatter in the $x$ and $y$ directions differs. Equation~(\ref{eq:bayes}) is also insensitive to rescaling the uncertainties of the data, because these enter in the same way both in the numerator and the denominator. According to \citet{jeffreys83}, $B > 10^{1/2}$ implies that the support for the fitted line is ``substantial'', and ``decisive'' if $B > 10^2$.

The employed likelihood model is only concerned with displacements of the observations perpendicular to the fitted line and provides no information on the distribution along the line \citep{hogg10}. To give a quantitative measure of the dynamic range of the data along the best-fit line, we project the data and their confidence ellipsoids on the best-fit line and calculate the weighted standard deviation of the data along the best-fit line $V$ and the median uncertainty along the best-fit line $W$. We define
\beq
R \equiv \frac{V}{W}.
\label{eq:r}
\eeq
In the case of one-dimensional data, $R$ is a measure of the intrinsic scatter, for example, $R=3$ would imply approximately $3\sigma$ significance of the intrinsic scatter. Here, the information provided by $R$ is complementary to the fitted slope and its uncertainty. Very large values of $R$ together with small relative uncertainty on the slope and high $B$ imply a strong correlation. If the confidence ellipsoids are oriented along the observed correlation and the off-diagonal elements of $\cf$ are ignored, $R$ will be artificially higher and the observed correlation will appear stronger. Unlike $B$, the parameter $R$ depends on the absolute values of data uncertainties.

\section{Intrinsic scatter in the $\lpl$, $\eexp$ and $\mni$ correlation}
\label{sec:scatter}

\begin{figure*}
\plottwo{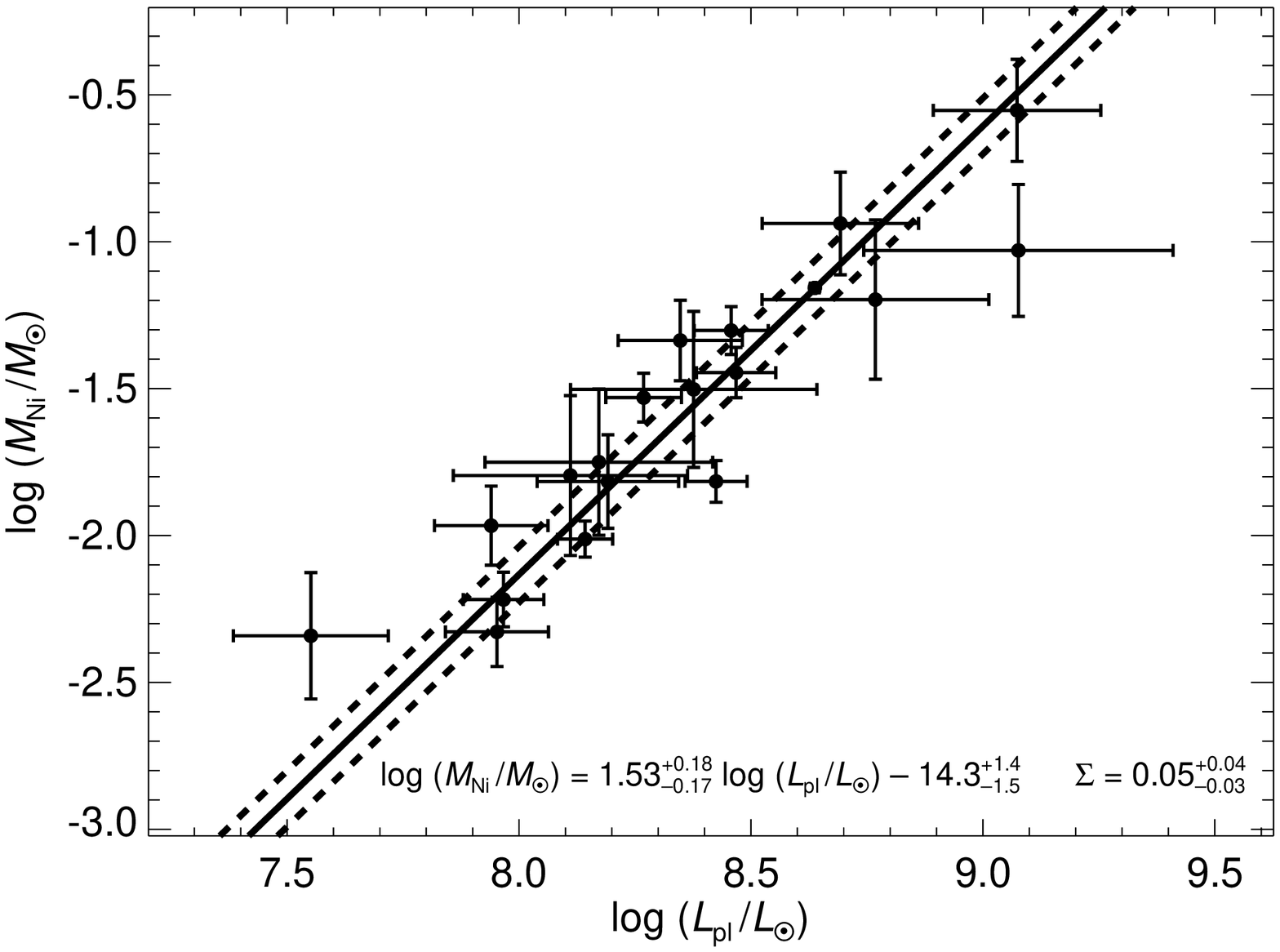}{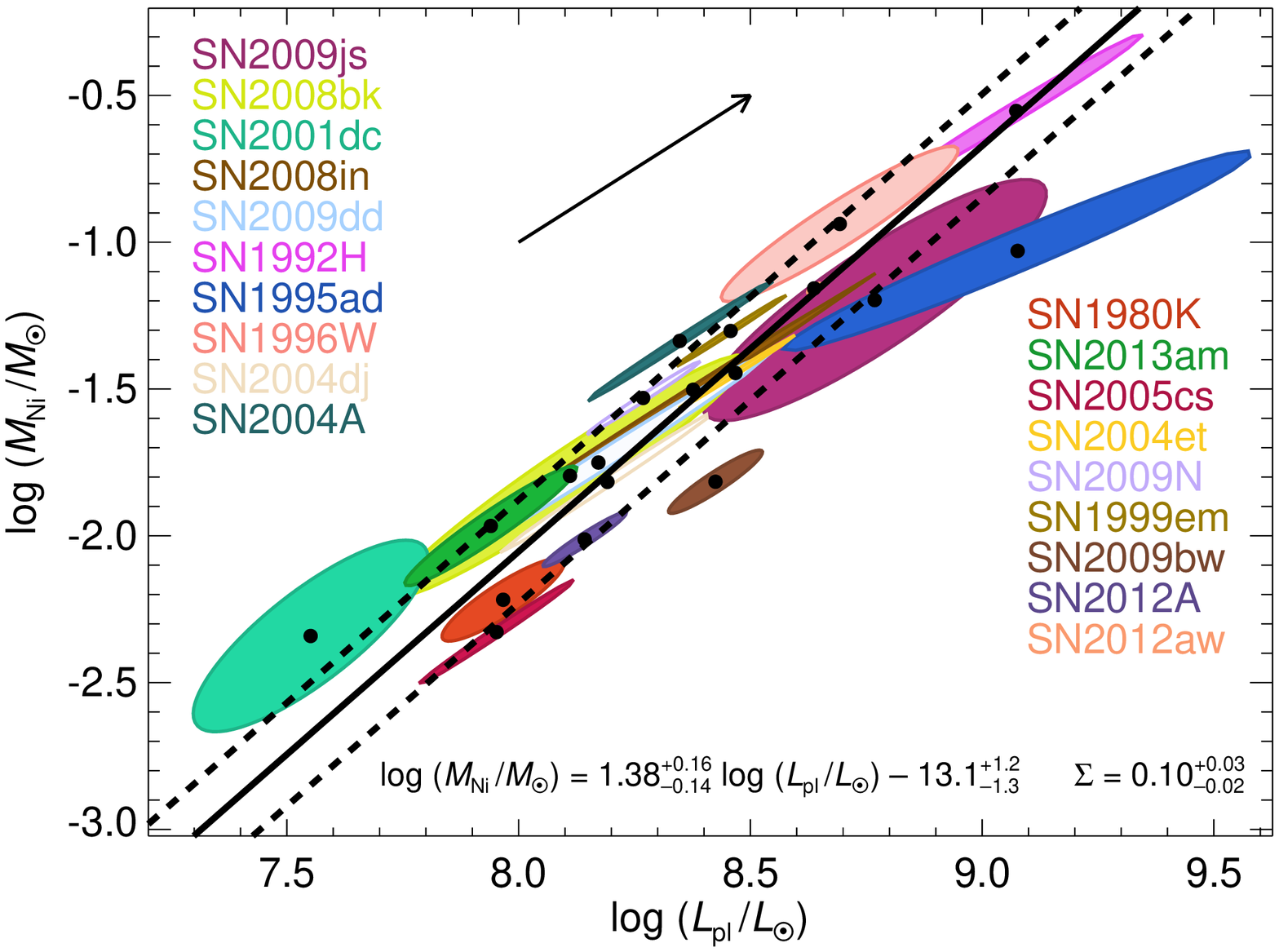}
\caption{The correlation between the plateau luminosity $\lpl$ (Eq.~[\ref{eq:lpl}]) and the nickel mass $\mni$ (Eq.~[\ref{eq:mni}]). The best-fit relation is shown with solid line and the intrinsic width $\Sigma$ is indicated with dashed lines. {\em Left}: The relation as usually presented, treating the parameter uncertainties as if they are uncorrelated.  We show one-dimensional uncertainty projections and the correlation does not exhibit internal spread. {\em Right}: The full confidence ellipsoids for each supernova imply a significant internal spread in the correlation. The arrow shows the direction of the covariance created by the distance uncertainty. }
\label{fig:mni}
\end{figure*}

Significant off-diagonal terms in $\cf$ occur when the uncertainties are dominated by a single systematic uncertainty $\sigma_{\rm syst}$, typically in the distance or the extinction\footnote{Our distance estimates are compared to previous results and other techniques (Cepheids, Type Ia supernovae) in \citetalias{pp14} and yield generally good agreement. Similarly good agreement is obtained for reddenings.}. Quantities linearly proportional to $\lbol$ such as $\lpl$ and $\mni$ are particularly susceptible. In other words, a bias in the distance will move $\lpl$ and $\mni$ in the same direction simultaneously by the same amount, introducing a covariance in these two parameters. Schematically, the covariance matrix of $\log (\lpl,\mni)$ is 
\beq
\mathbf{C}^{\log \lpl, \log\mni} = \left(
\begin{array}{cc}
\sigma_{{\rm pl}}^2 + \sigma_{\rm syst}^2 & \sigma_{\rm syst}^2 \\                        
\sigma_{\rm syst}^2 & \sigma_{{\rm tail}}^2 + \sigma_{\rm syst}^2
\end{array}
\right),
\label{eq:c2}
\eeq
where $\sigma_{\rm pl}$ and $\sigma_{\rm tail}$ are the uncertainties in the observed magnitudes during the plateau and the exponential decay tail, respectively. Usually, $\sigma_{\rm pl}, \sigma_{\rm tail} \ll \sigma_{\rm syst}$ and the confidence ellipsoid is strongly elongated in the direction of the systematic uncertainty. If $\log \lpl$ and $\log \mni$ are perfectly correlated, neglecting the off-diagonal terms in Equation~(\ref{eq:c2}) will imply a value of $R$ that is twice the true value. Even if the full covariance matrix is not available, an approximate covariance matrix similar to Equation~(\ref{eq:c2}) can be constructed to properly visualize the confidence ellipsoids.

In the left panel of Figure~\ref{fig:mni}, we show the estimates of $\lpl$ and $\mni$ with uncertainties simply represented by the diagonal terms of $\cf$, as is commonly done \citep[e.g.][]{hamuy03,anderson14,spiro14}. We would infer that there is a linear correlation between $\log\lpl$ and $\log\mni$ with a slope of $1.51^{+0.17}_{-0.17}$ and $R=4.2$, implying a strong correlation. More importantly, considering only the diagonal terms gives a false impression that all the points are compatible with the best-fit line given their uncertainties. Quantitatively, there is no evidence for intrinsic scatter, with $\Sigma = 0.05^{+0.04}_{-0.03}$. 

When the confidence ellipsoids are properly included as in the right panel of Figure~\ref{fig:mni}, the picture changes. The confidence ellipsoids are significantly elongated, because of their mutual dependence on distances, as indicated by the arrow\footnote{Note that the uncertainties in absolute magnitude and expansion velocity \citep[e.g.][]{poznanski13} should not be very correlated, unless the velocities were used for an estimate of the distance modulus, in which case there should be a significant correlation.}. The correlation is noticeably less significant, $R=3.0$, although there is no doubt this correlation exists given the large dynamic range of the parameters. The Bayes factor is $B \approx 9\times 10^7$ implying strong support for the correlation. More importantly, we discover a statistically significant intrinsic width of the relation $\Sigma = 0.12^{+0.03}_{-0.02}$, which implies a scatter of $0.2$\,dex in $\mni$ for a fixed $\lpl$. Furthermore, neglecting the off-diagonal terms or the intrinsic width of the relation can bias the inferred slope \citep[e.g.][]{tremaine02}. Neglecting the off-diagonal terms increases the slope by about $0.19$ with a corresponding change in the intercept. Not accounting for the intrinsic scatter leads to slopes of $1.83^{+0.07}_{-0.06}$ and $1.56^{+0.12}_{-0.11}$ for the full and diagonal covariance matrix, respectively. 

\begin{figure*}
\plottwo{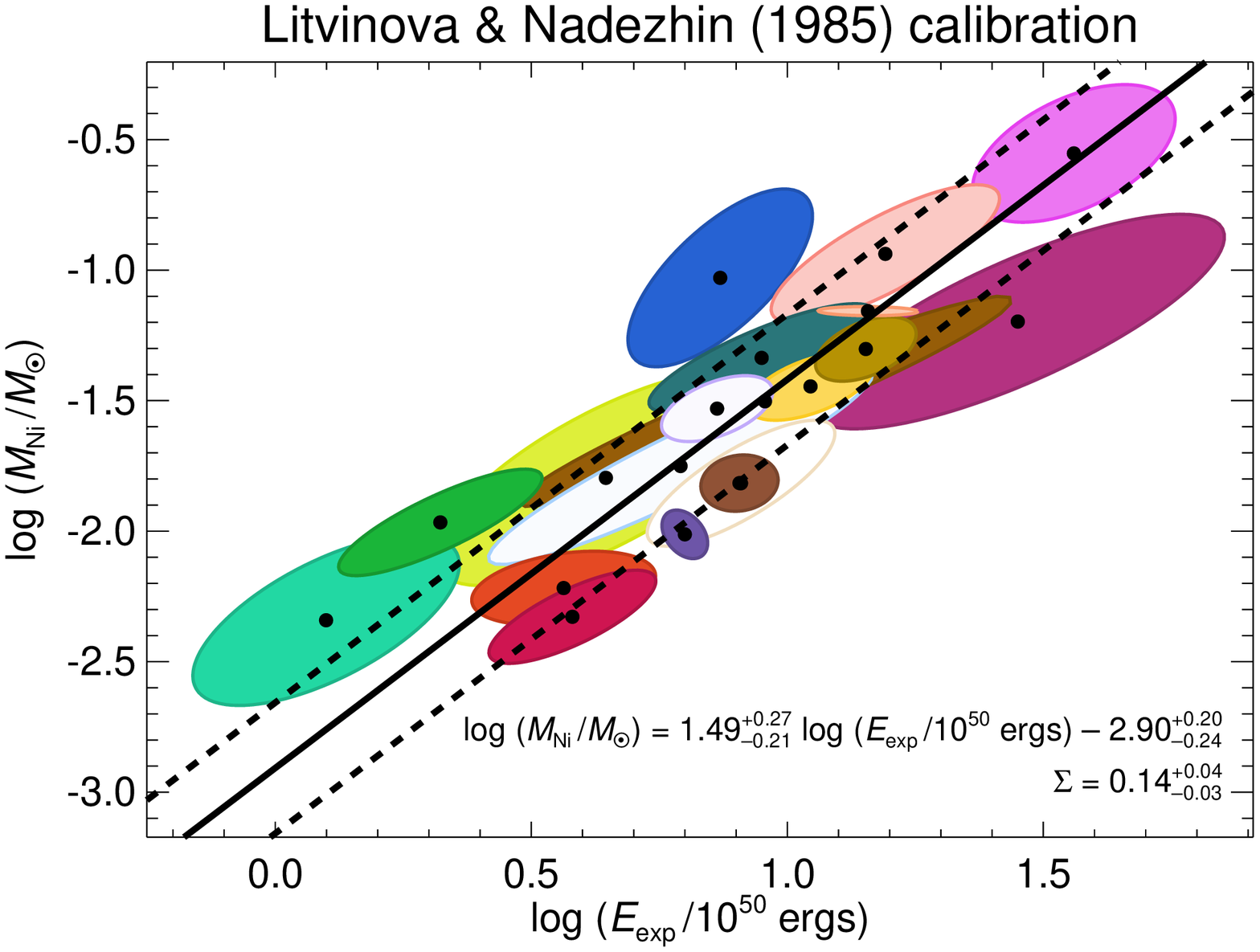}{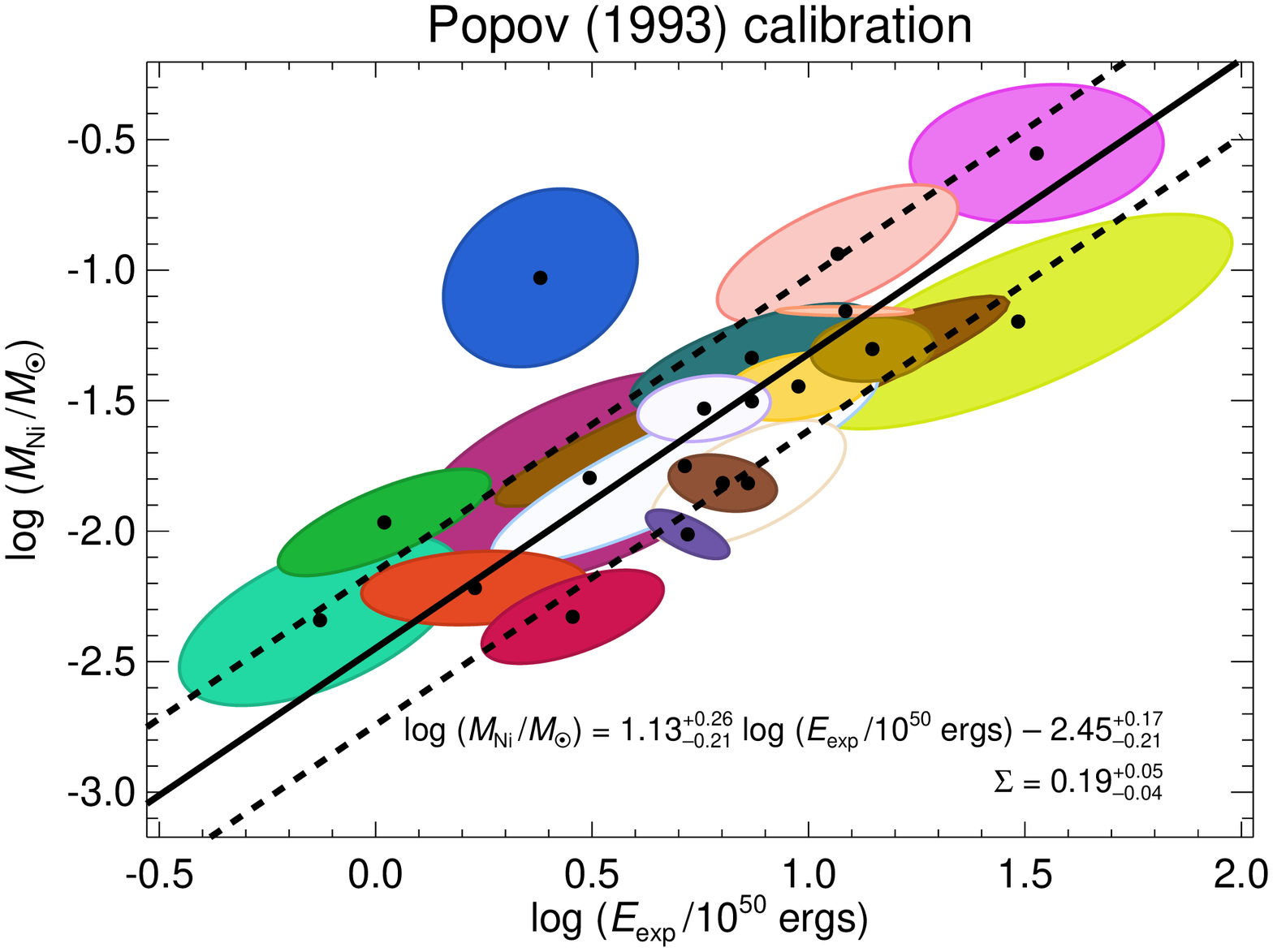}
\caption{Nickel mass $\mni$ as a function of explosion energy $\eexp$, with the confidence ellipses properly visualized. The colors of the individual supernovae are the same as in the right panel of Figure~\ref{fig:mni}. We use the scaling relations of \citet[left panel]{litvinova85} and \citet[right panel]{popov93}.}
\label{fig:e_mni}
\end{figure*}

Since $\lpl$ does not have an immediate physical interpretation, we show $\mni$ as a function of $\eexp$ in Figure~\ref{fig:e_mni} for the scaling relations of \citet{litvinova85} and \citet{popov93}. The relative position of the majority of the data points remains unchanged when compared to the right panel of Figure~\ref{fig:mni}, which indicates that $\lpl$ is a good proxy for $\eexp$. There are small differences between the two scaling relations, but the relative positions of the majority of the points are unchanged. For the \citet{litvinova85} coefficients, we find that the $\eexp$--$\mni$ correlation is less significant than $\lpl$--$\mni$ with $R=2.8$ and $3.7$ for the full and diagonal covariance matrix, respectively. The Bayes factor is $B \approx 1.7 \times 10^5$ indicating strong correlation, but weaker than $\lpl$--$\mni$. The inferred intrinsic width orthogonal to the line is slightly higher than in the $\lpl$--$\mni$ correlation but again statistically significant, $\Sigma = 0.14^{+0.04}_{-0.03}$ or $0.25$\,dex in $\mni$ for fixed $\eexp$. The intrinsic width from the \citet{popov93} calibration is $\Sigma = 0.19^{+0.05}_{-0.04}$.

\begin{figure*}
\plottwo{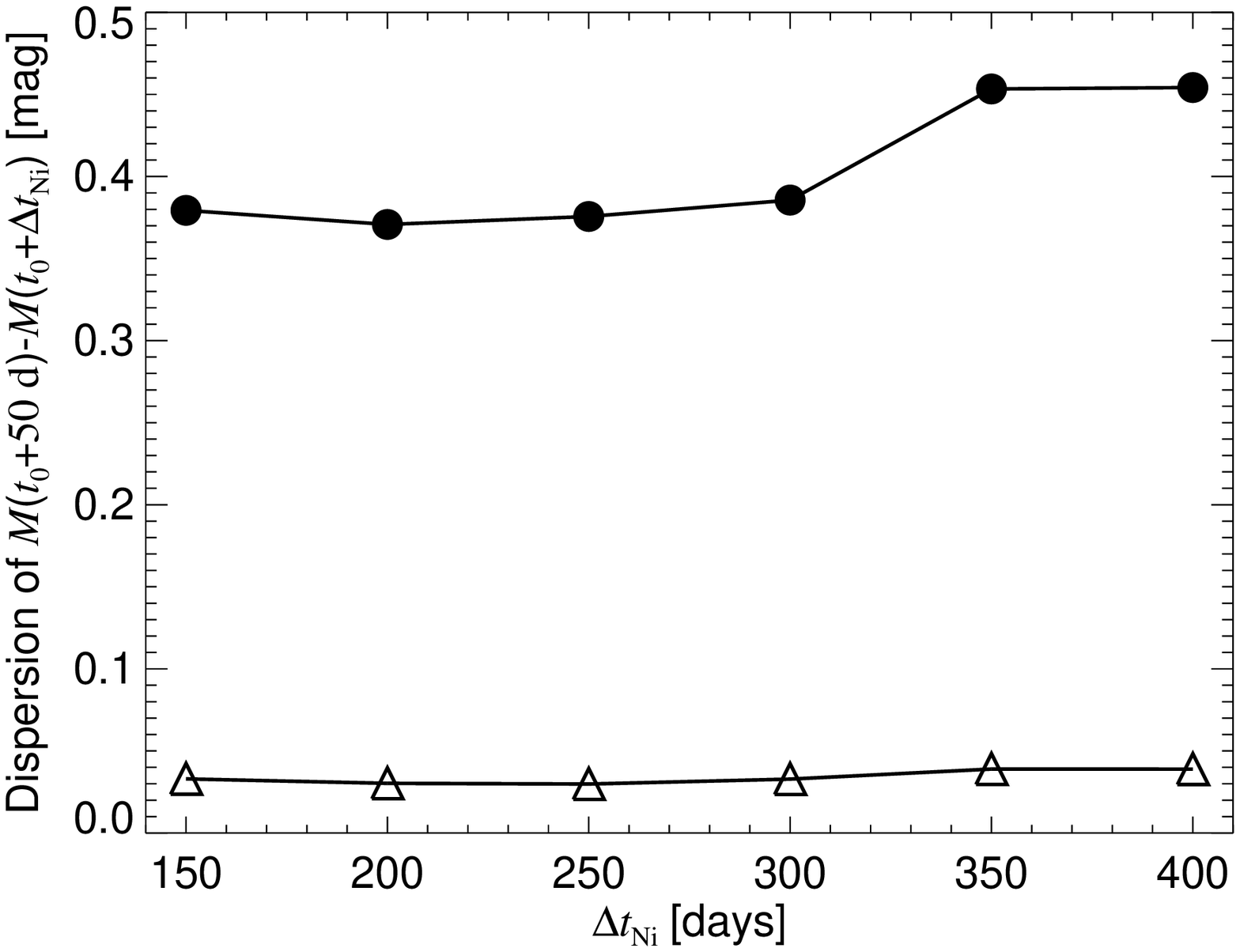}{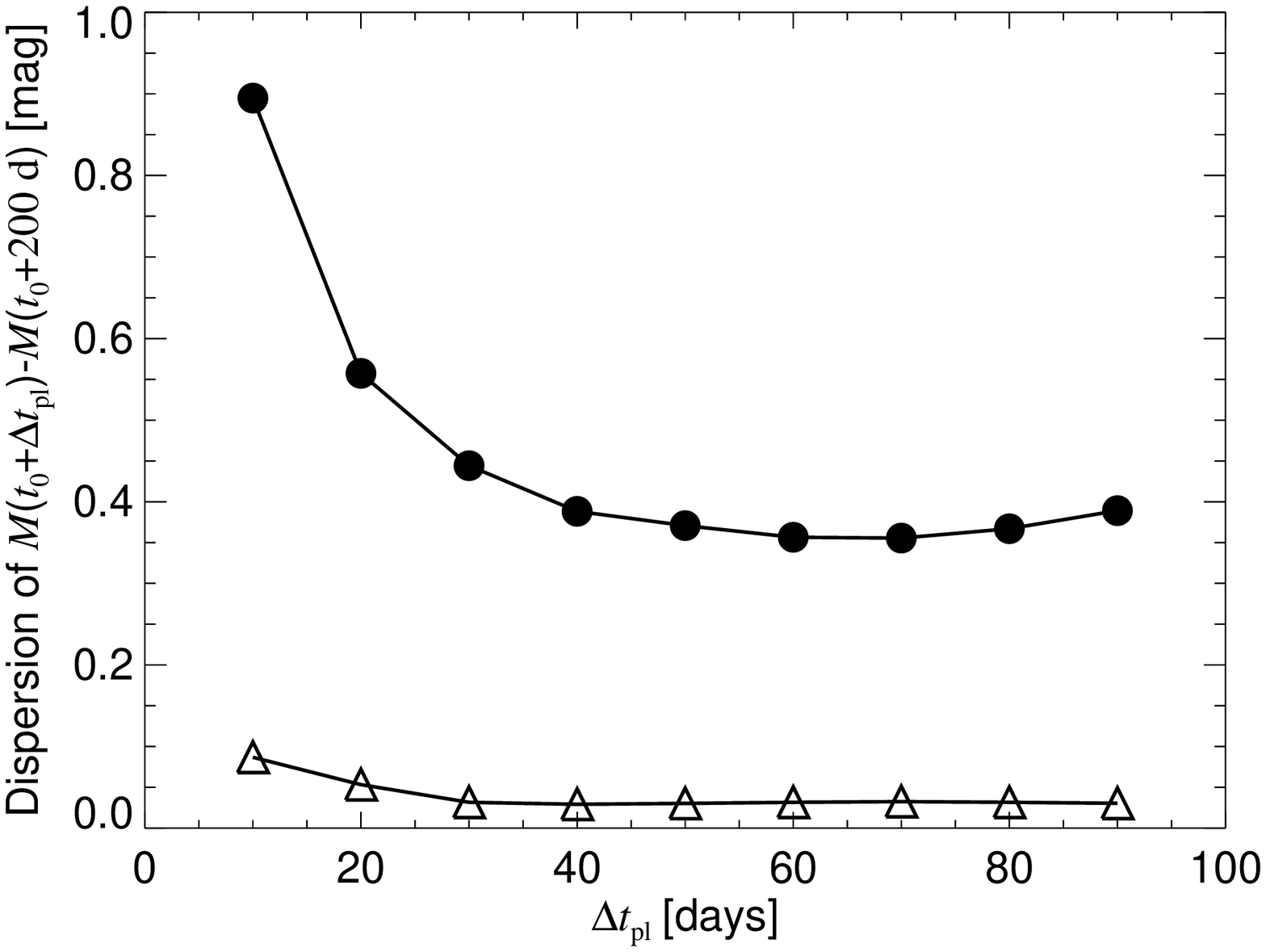}
\caption{Weighted standard deviation of the bolometric magnitude difference between the plateau and the exponential tail are shown with filled circles as function of $\tni$ (left panel) or $\tpl$ (right panel). There is little dependence on $\tni$ indicating that the different $\gamma$-ray trapping efficiencies are not responsible for the intrinsic width seen in Figures~\ref{fig:mni} and \ref{fig:e_mni}. In other words, the exponential decay is nearly parallel for stars in our sample. Open triangles indicate the mean uncertainty of the individual measurements used. Only supernovae with data spanning $\tpl$ and $\tni$ are used.}
\label{fig:diff}
\end{figure*}

The intrinsic width of the $\eexp$--$\mni$ correlation could be due to the $\gamma$-ray trapping efficiencies $\ag$ varying among supernovae with the same $\eexp$. Since the exponential decay luminosity is proportional to $[1-\exp(-\ag/\tni^2)] \exp (-\tni/\tau)$ \citep[e.g.][]{chatzopoulos12,nagy14}, where $\tni$ is the time elapsed since the explosion, supernovae with significant $\gamma$-ray leakage not only appear fainter at any point of this phase but also decay faster \citep{anderson14}, and their light curves diverge from those of supernovae with full $\gamma$-ray trapping over time. Since our sample contains supernovae with decay rates compatible with full trapping as evidenced by the exponential decay slopes \citepalias{pp14}, the $0.25$\,dex difference in the inferred $\mni$ at $\tni = 200$\,days should increase to about $0.7$\,dex at $\tni=400$\,d, if the scatter is due to $\gamma$-ray leakage in some objects. 

To test whether the late light curves of supernovae in our sample diverge with time due to incomplete $\gamma$-ray trapping in some objects, we show the weighted standard deviation of the bolometric magnitude difference between the plateau and the exponential tail as a function of the time elapsed since explosion, $\tni$, in the left panel of Figure~\ref{fig:diff}. We see that the bolometric magnitude dispersion increases from $0.37$\,mag at $\tni=200$\,d to $0.45$\,mag at $\tni=400$\,days, much less than what we would expect if some supernovae showed full trapping and some only partial. This means that the slopes of the exponential decay are very similar among our objects and are compatible with full $\gamma$-ray trapping.

For the sake of completeness, we test what is the importance of when is the plateau luminosity determined. We show the weighted standard deviation of the bolometric magnitude difference between the plateau and the exponential tail but now as a function of $\tpl$ in the right panel of Figure~\ref{fig:diff}. For small $\tpl$, the dispersion is relatively high, presumably due to differences in the properties of the shock-heated ejecta shortly after shock breakout, but for $\tpl \gtrsim 40$\,d the dispersion remains approximately constant. We conclude that the intrinsic width of the $\eexp$--$\mni$ correlation is robust with respect to when exactly the plateau and exponential decay tail luminosities are measured, and that it is not due to variations in the $\gamma$-ray leakage\footnote{An additional piece of anecdotal evidence against significant $\gamma$-ray leakage comes from comparing SN2013am and SN2005cs, which have nearly identical luminosities for the first $\sim 70$\,days. However, SN2013am has a noticeably shorter $\tp$ and a higher inferred $\mni$ than SN2005cs. This implies that SN2013am has slightly smaller $\eexp$ and significantly smaller $\mej$ than SN2005cs (Fig.~\ref{fig:m_e}). If $\gamma$-ray escape were important, we would expect smaller inferred $\mni$ and faster exponential decay in SN2013am than in SN2005cs. Yet, the exponential decay slope is almost the same in both objects \citep{zhang14} and SN2013am has higher inferred $\mni$.}.

\section{The correlation between $\mej$ and $\eexp$}
\label{sec:m_eexp}

\begin{figure*}
\plottwo{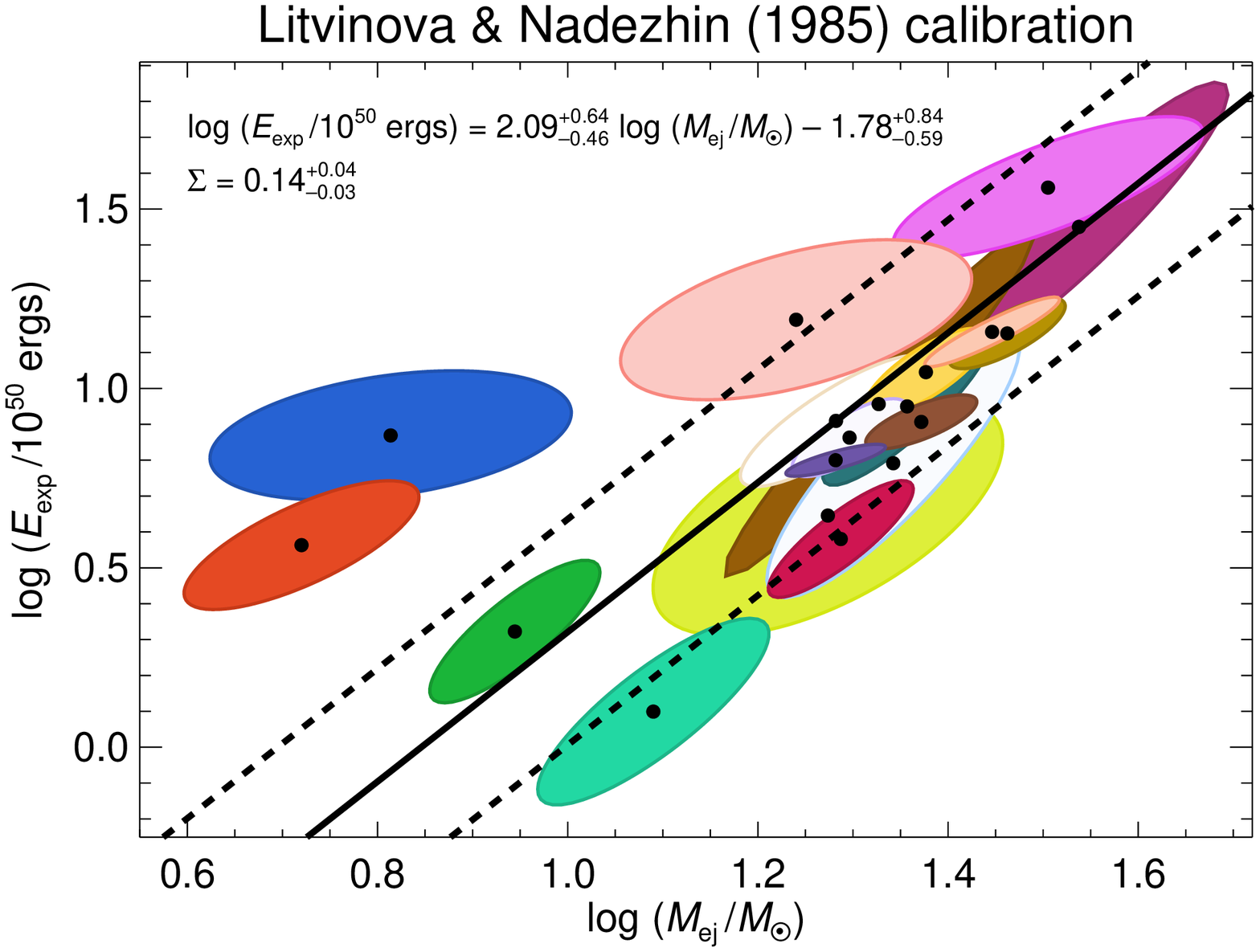}{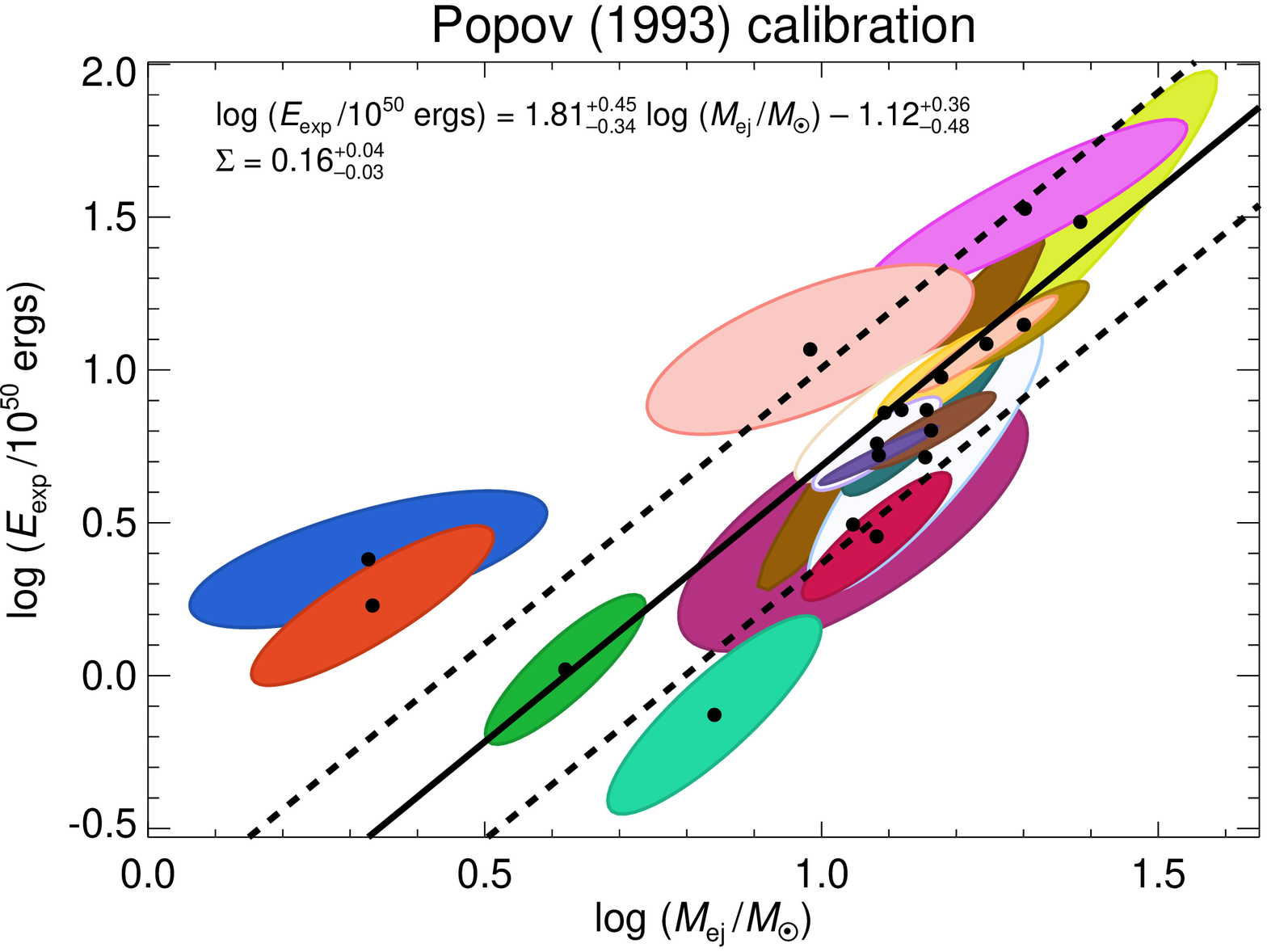}
\caption{Explosion energy $\eexp$ as a function of ejecta mass $\mej$. The colors of the individual supernovae are the same as in the right panel of Figure~\ref{fig:mni}. We use the scaling relations of \citet[left panel]{litvinova85} and \citet[right panel]{popov93}.}
\label{fig:m_e}
\end{figure*}

Correlated uncertainties in $\mathbf{f} =  \log (\lpl, \mni, \eexp, \mej)$ can also occur when some of the vectors $\partial f_i/\partial \mathbf{b}$ are nearly parallel. Equations~(\ref{eq:eexp}--\ref{eq:mej}) imply that the covariance of $\log \mej$ and $\log \eexp$ is
\beq
\frac{C_{\log\mej,\log\eexp}}{\sqrt{C_{\log\mej,\log\mej}C_{\log\eexp,\log\eexp}}} \sim \frac{\bm{\alpha}\cdot\bm{\beta}}{|\bm{\alpha}||\bm{\beta}|},
\label{eq:offdiag}
\eeq
if the uncertainties in the individual components of $\mathbf{b}$ are approximately the same. This is reasonable, because $5$ to $10\%$ uncertainties in the distance modulus, $\tp$ and $v$ are expected. We obtain a high correlation of $0.94$ and $0.97$ for the $\bm{\alpha}$ and $\bm{\beta}$ from \citet{litvinova85} and \citet{popov93}, respectively. This degeneracy comes the physics of the Type II-Plateau supernova light curves, where higher kinetic energy makes the material transparent earlier, which can be compensated for by increasing ejecta mass to produce approximately the same plateau duration and luminosity \citep[e.g.][]{arnett80,litvinova85,kasen09,dessart10,dessart13}. The degeneracies between parameters describing supernova light curves were also investigated by \citet{nagy14} using semi-analytic models. As a result, inferences of $\eexp$ and $\mej$ will be highly correlated in any technique based on light curves and expansion velocities.


In Figure~\ref{fig:m_e} we show $\mej$ and $\eexp$ as inferred for our sample and the two scaling relations. The uncertainty ellipsoids are elongated along the correlation, as expected. We find significant systematic offsets between the scaling relationships, in particular \citet{popov93} calibration produces smaller and more realistic $\mej$, and smaller $\eexp$. The relative positions of the individual supernovae remains unchanged in most cases. For the \citet{litvinova85} calibration, we find a relatively high uncertainty for the slope $2.09^{+0.64}_{-0.46}$ and a statistically significant intrinsic width to the correlation, $\Sigma = 0.14^{+0.04}_{-0.03}$. We find $R=2.0$ and the Bayes factor of $B\approx 200$, which implies that this correlation is much weaker than $\eexp$--$\mni$, albeit the evidence is still ``decisive'' in the classification of \citet{jeffreys83}. The results are similar for the \citet{popov93} calibration. 

There are two outliers to the $\mej$--$\eexp$ correlation: SN1995ad and SN1980K. The Type II-Linear SN1980K has well-constrained distance \citepalias{pp14} and relatively good light curve covering the transition to the exponential decay. Recent investigations of large samples of hydrogen-rich supernovae indicate that there is a continuum of light curve shapes between Type II-Plateau and Type II-Linear (\citealt{anderson14} and \citealt{sanders15}, but see also \citealt{arcavi12} and \citealt{faran14}), suggesting that a larger unbiased sample of supernovae might fill the space between SN1980K and the rest of our sample. 

SN1995ad has distinctively shorter $\tp \approx 85$\,days than most supernovae in our sample, as found also by \citet{inserra13}, which is responsible for the outlying results. The explosion epoch of \citet{inserra13} is about $12$\,days later than ours and their adopted distance modulus is about $0.3$\,mag closer (although the distances are compatible within their $1\sigma$ limits) than what we obtained in \citetalias{pp14}. From radiation-hydrodynamical modeling \citet{inserra13} found $\log(\eexp, \mej) \approx (0.3, 0.7)$ in the units of Equations~(\ref{eq:eexp}--\ref{eq:mej}), while we find $(0.4, 0.3)$ and $(0.9, 0.8)$ using the relations of \citet{popov93} and \citet{litvinova85}, respectively. Without the outlying SN1995ad and SN1980K, the $\mej$--$\eexp$ correlation exhibits a steeper slope of $2.60^{+0.41}_{-0.30}$ \citep[compatible with the result of][]{poznanski13}, significantly smaller intrinsic width $\Sigma=0.05^{+0.02}_{-0.01}$, and $R=1.9$. The Bayes factor increases to $B \approx 1.8 \times 10^4$, as expected when outliers are removed.


\begin{figure*}
\plotone{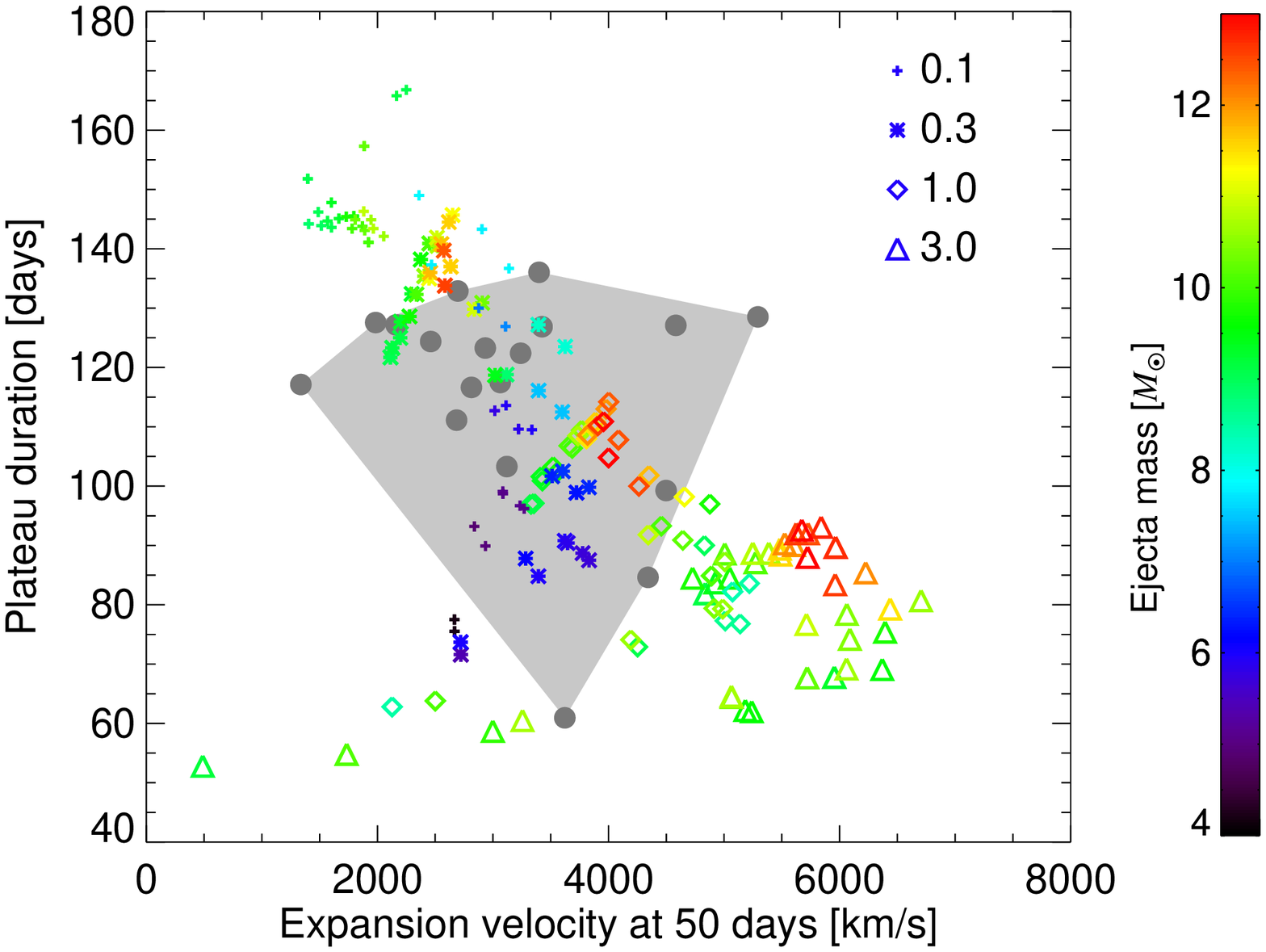}
\caption{Plateau durations and expansion velocities from the theoretical models and the observations. The results of the theoretical non-rotating progenitor explosions of \citet{dessart10} are shown with plus signs, stars, diamonds and triangles for explosion energies of $0.1$, $0.3$, $1.0$ and $3.0 \times 10^{51}$\,ergs. The color of the symbols indicate the ejecta mass. The observations are shown with gray solid circles and their convex hull is marked with the gray polygon. The typical parameters of Type II-Plateau supernovae ($\tp \approx 120$\,days, $v \approx 3500$\,km\,s$^{-1}$) can be explained by a range of models going from low $\eexp$ and low $\mej$ to moderate $\eexp$ and high $\mej$. Finer details of the light curves and spectra need to be analyzed to break this degeneracy.}
\label{fig:vej_tpl}
\end{figure*}

Since we are not interested in the specific values of $\eexp$ and $\mej$, but rather in the process of their estimation, the results of the analytic scaling relations of Equations~(\ref{eq:eexp}--\ref{eq:mej}) are only approximate and must be confronted with more detailed models. In Figure~\ref{fig:vej_tpl} we show $\tp$ and $v$ of the radiation-hydrodynamic models of explosions of non-rotating red supergiant progenitors of \citet{dessart10} with information on $\eexp$ and $\mej$ encoded in symbols and their colors. The models do not include heating by radioactive nickel and thus underestimate $\tp$. The coverage of the parameter space by the theoretical models is not uniform, and in particular some observed supernovae fall in the areas where there are no models, making interpolation in these theoretical models to estimate the supernova parameters difficult. More importantly, the bulk of the observed supernovae cluster at $\tp \approx 120$\,days and $v \approx 3500$\,km\,s$^{-1}$, and these parameters can be explained by a range of models going from low explosion energy ($\eexp = 0.1 \times 10^{51}$\,ergs) and low ejecta mass ($\mej \approx 6\,\msun$) to normal explosion energy ($\eexp = 1.0\times 10^{51}$\,ergs) and relatively high ejecta mass ($\mej \gtrsim 10\,\msun$). In other words, very different sets of theoretical parameters will yield very similar observables, at least for $\tp$ and $v$ (and likely also $\lpl$ due to the observed $\lpl$--$v$ correlation), implying that also the more sophisticated models are prone to the same degeneracy as the linear scaling relations.
\\

\section{Discussions and conclusions}
\label{sec:disc}

We show that the correlated uncertainties between parameters derived from supernova light curves and velocities are strong and are oriented along the parameter correlations. The covariances between the quantities arise either due to uncertainties in the distance affecting two quantities in the same way ($\lpl$ and $\mni$, Fig.~\ref{fig:mni}), or due to degeneracies inherent to the physics of the supernova light curves ($\eexp$ and $\mej$, Fig.~\ref{fig:m_e}). As a result, the statistical significance of these correlations is reduced, but the correlations cannot be fully explained away by the covariances. The correlation of $\mej$ and $\eexp$ is weaker than the other two correlations we investigated ($\lpl$, $\eexp$, and $\mni$) and its significance is sensitive to whether SN1980K (Type II-Linear) and SN1995ad (Type II-Plateau with short plateau duration) are included. Conversely, properly characterizing the uncertainty ellipsoids reveals an intrinsic width to the $\lpl$--$\mni$ and $\eexp$--$\mni$ correlations (Figs.~\ref{fig:mni} and \ref{fig:e_mni}). Now we explore the astrophysical implications of these findings.

In studies of the neutrino mechanism, most of $\eexp$ comes from the neutrino-driven wind emanating from the nascent proto-neutron star \citep[e.g.][]{scheck06,ugliano12,pt14}. In a simple spherical picture, the evolution of the neutrino-driven wind is determined by the thermodynamic structure of the layers below the ejecta mass cut \citep{pt14}, while the ejected mass of \nif\ depends primarily on the mass of the shock-heated ejecta exposed to sufficiently high temperatures \citep{weaver80,woosley88,thielemann90}. The intrinsic scatter in the $\eexp$--$\mni$ relation therefore implies that the progenitor structure below and above the mass cut cannot be fully described by a single parameter, such as the compactness \citep{oconnor11,oconnor13,nakamura14,pt14,perego15}. Using the results of \citet{pt14}, we find a width of $0.10^{+0.01}_{-0.01}$\,dex in their $\eexp$--$\mni$ correlation, which is consistent with the results presented here. The theoretical predictions of the correlation slope \citep{pt14} approximately agree with the observations presented here, but disagree with the inferences presented by \citet{hamuy03}. A similar conclusion on the presence of multiple parameters determining the properties of core-collapse supernova light curves was reached by \citet{sanders15}, who quantified a dispersion in the decline rate-peak magnitude relation of hydrogen-rich supernovae.

Changes in the intrinsic width of the $\eexp$--$\mni$ correlation as a function of $\eexp$ or other parameters can further constrain the explosion physics. For example, the spread in $\mni$ at constant $\eexp$ could be caused by a varying amount of \nif\ fallback in different progenitors. We would expect that the fallback will be generally less important at higher $\eexp$ and the spread of $\mni$ should thus increase as $\eexp$ decreases. In principle, this is testable given a large set of well-observed supernova explosions. New unbiased surveys of bright nearby supernovae such as ASAS-SN \citep{holoien14,shappee14} are particularly useful due to feasibility of detailed follow-up observations and the exploration of new parts of the parameter space (e.g., low-metallicity stellar environments).

Contrary to the common picture \citep[e.g.][]{heger03,nomoto06,utrobin09,utrobin14}, there is little evidence from the parameterized studies of the neutrino mechanism that the supernova properties such as $\eexp$ or $\mni$ will strongly correlate with the mass of the progenitor \citep[e.g.][]{oconnor11,ugliano12,bruenn14,nakamura14,pt14,perego15,ertl15}, because the ultimate fate of the star and the initiation of the explosion is set by the physics and the thermodynamic structure on the inner $\sim 2.5\,\msun$ of the progenitor, which is not monotonic with the initial mass, metallicity or final hydrogen mass \citep{sukhbold14}.

The degeneracy between $\mej$ and $\eexp$ can be reduced by modeling finer features in the light curve and spectra, such as the O I 6303-6363\,\AA\ line advocated by \citet{dessart10}. However, characterizing these features generally requires substantial observational effort on nearby objects (for example the O I 6303-6363\,\AA\ line fully develops only $\sim 300$\,days after explosion; \citealt{dessart10}), along with confidence in the underlying physics and considerable effort in its numerical implementation. Surveys focusing on bright, nearby supernovae such as ASAS-SN \citep{holoien14,shappee14} could be of help in this regard. It will be much harder to obtain such detailed information in the future dominated by primarily photometric discovery machines such as the Large Synoptic Survey Telescope \citep[e.g.][]{ivezic14}. As a result, reliable recovery of supernova parameters from light curves and (potentially scarce) expansion velocities will only grow in importance, mandating rigorous uncertainty analysis that consistently include all relevant contributions to the final uncertainty budget, as we have done here. Detailed investigations of the supernova parameter covariances, degeneracies and multiple solutions based on detailed radiation hydrodynamic models is the next logical step in the preparation for the upcoming surveys. The understanding of the intricacies of the parameter recovery will not only yield greater physical understanding of the supernova population, but can influence back the design and strategy of the surveys currently in preparation.

\section*{Acknowledgements}
We thank Jujia Zhang for providing us with the velocities of SN2013am. We thank Chris Kochanek, Michael Strauss, Bruce Draine, and Todd Thompson for a detailed reading of the manuscript. We thank the referee for comments. OP appreciates the discussions with Adam Burrows. We thank the referee for comments that helped to improve the manuscript. Support for OP was provided by NASA through Hubble Fellowship grant HST-HF-51327.01-A awarded by the Space Telescope Science Institute, which is operated by the Association of Universities for Research in Astronomy, Inc., for NASA, under contract NAS 5-26555. Support for JLP is provided in part by the Ministry of Economy, Development, and Tourism's Millennium Science Initiative through grant IC120009, awarded to The Millennium Institute of Astrophysics, MAS.

\end{document}